\documentclass[12pt,final,notitlepage]{iopart}
\usepackage{graphicx}
\usepackage{bm}
\usepackage{iopams}
\usepackage{psfrag}
\usepackage{showkeys}

\newcommand{\ten}[1]{\mbox{\textbf{\textsf{#1}}}}
\newcommand{\mi}{\mathrm{i}}

\begin{document}

\article{CEWQO 2008}{Resonant Casimir--Polder forces in planar
meta-materials}

\author{Agnes Sambale$^1$, Stefan Yoshi Buhmann$^2$, Ho Trung Dung$^3$
and Dirk-Gunnar Welsch$^1$}

\address{1 Theoretisch--Physikalisches Institut,
Friedrich--Schiller--Universit\"at Jena,
Max--Wien--Platz 1, D-07743 Jena, Germany}

\address{2 Quantum Optics and Laser Science, Blackett Laboratory,
Imperial College London, Prince Consort Road,
London SW7 2BW, United Kingdom}

\address{3 Institute of Physics, Academy of
Sciences and Technology, 1 Mac Dinh Chi Street,
District 1, Ho Chi Minh city, Vietnam}

\ead{agnes.sambale@uni-jena.de}

\date{\today}

\begin{abstract}
We study the resonant Casimir--Polder potential of an excited atom
near a half space containing magneto-electric meta-material of
various kinds on the basis of macroscopic quantum electrodynamics.
Analytical results are obtained in the nonretarded and retarded
distance regimes and numerical examples are given. We compare our
findings with the potential of an excited atom near a left-handed
superlens. 
\end{abstract}



\section{Introduction}

In view of the rapid progress made in the fabrication and application
of meta-materials \cite{Rockstuhl2007}, which often exhibit
left-handed properties \cite{Smith2000, Lezec2007}, a theoretical
understanding of the striking optical effects associated with such
materials is desirable. Mainly in the past decade, phenomena such as
negative refraction \cite{Veselago1968, Smith2000}, superlenses
\cite{Pendry2000,Grbic2004}, invisibility devices 
\cite{Leonhardt2006,Schurig2006} and the reversed Doppler effect
\cite{Grbic2002, Kats2007} have been investigated and measured. Of
particular interest to the field of nano-technologies is the research
on dispersion forces in the presence of meta-materials with specially
tailored magneto-electric properties described by complex
$\varepsilon(\omega)$ and $\mu(\omega)$. For example, the Casimir
force between two magneto-electric plates \cite{Henkel2005,Tomas2005},
as well as that between anisotropic meta-materials \cite{Rosa2008}
have recently been addressed where particular attention was drawn to
the question of whether attractive behavior can be turned into 
repulsion, an issue closely related to problem of stiction
\cite{Rosa2008a}. Whereas precision measurements of Casimir forces
between ordinary materials have already carried out \cite{Decca2003}
an experimental confirmation of dispersion interactions in
meta-materials remains to be done as a prospective task.  

We have recently investigated the resonant Casimir--Polder (CP)
interaction of an excited atom near a superlens of left-handed
material \cite{Sambale2008a}, finding that for very weak absorption
the CP potential may be enhanced near the focal plane of the lens.
In this article we study the resonant CP interaction near a
magneto-electric half space (Sec.~\ref{rvdW}). Endowed with an
understanding of the behavior of the potential near a single
interface, we then reexamine the superlens geometry in order
to clarify whether focal-plane enhancement is a genuine left-handed
effect or whether similar behavior can be observed with ordinary
magnetoelectrics (Sec.~\ref{plg}). 


\section{Magneto-electric half space}
\label{rvdW}

It has been recognized that the impact of the magneto-electric
properties of the meta-material (e.g.~left-handedness) on dispersive
interactions is much more pronounced for excited systems
due to fact that selected frequencies (i.e., atomic transition
frequencies) contribute dominantly to the electromagnetic response.
For an excited atom [position $\mathbf{r}_A$, energy eigenstate
$|n\rangle$, transition frequencies $\omega_{nk}$, electric-dipole
transition matrix elements $\mathbf{d}_{nk}$] the off-resonant part of
the interaction can often be neglected, so that the CP potential may
be given as \cite{Buhmann2004, Review2006}
\begin{eqnarray}
\label{equ1}
U_n
(\mathbf{r}_A)=\nonumber\\
-\mu_0\sum_{k<n}
\omega_{nk}^2\mathbf{d}_{nk}\!\cdot\!
\mathrm{Re}\ten{G}^{(1)}(\mathbf{r}_A,\mathbf{r}_A,\omega_{nk})
\!\cdot\!\mathbf{d}_{kn}
\end{eqnarray}
($\Theta$: unit step function, $\ten{G}^{(1)}$: scattering Green
tensor).

Consider first an atom placed in front of a magneto-electric half
space (at $z_A\le 0$) of permittivity $\varepsilon(\omega)$ and
permeability $\mu(\omega)$. The associated scattering Green tensor
reads (cf., e.g.~\cite{Kampf2005})
\begin{eqnarray}
\label{Green}
\ten{G}^{(1)}(z_A,z_A,\omega_{nk})\\
=\frac{\mi}{8\pi}\int _0^\infty {\rm d}q\,
\frac{q}{\beta}\,e^{2\mi\beta z_A}
\Biggl[\Biggl(r_s-\frac{\beta^2c^2}{\omega_{nk}^2}\,
r_p\Biggr)({\bf e}_x{\bf e}_x+{\bf e}_y{\bf e}_y)\nonumber\\
+2\,\frac{q^2c^2}{\omega_{nk}^2}\,r_p
{\bf e}_z{\bf e}_z\Biggr]\nonumber
\end{eqnarray}
where
\begin{eqnarray}
\label{r21}
r_s=\frac{\mu\beta-\beta_{1}}{\mu\beta+\beta_{1}}\,,&\;&
r_p=\frac{\varepsilon\beta-\beta_{1}}
 {\varepsilon\beta+\beta_{1}},\\
\label{beta}
\beta=\sqrt{\frac{\omega_{nk}^2}{c^2}-q^2}\,,&\;&
\beta_1=\sqrt{\varepsilon\mu\frac{\omega_{nk}^2}{c^2}-q^2}\,.
\end{eqnarray}
[$\varepsilon\equiv \varepsilon(\omega_{nk})$, $\mu \equiv
\mu(\omega_{nk})$]. The square root of $\beta_1$ has to be chosen such
that $\mathrm{Im}\beta_1>0$ for a passive medium. Note that when
the atom is embedded in a medium, local-field corrections need to be
taken into account \cite{Sambale2007}. 

Substituting (\ref{Green}) into (\ref{equ1}) gives for the resonant CP
potential
\begin{eqnarray}
\label{Ur2}
 U_n(z_A)=\nonumber\\
 -\mu_0\sum_{k<n}\omega_{nk}^2
 [\mathrm{Re}G^{(1)}_{xx}(z_A,z_A,\omega_{nk})
 |\mathbf{d}_{nk}^\parallel|^2\nonumber\\
+\mathrm{Re}G^{(1)}_{zz}(\mathbf{r}_A,\mathbf{r}_A,\omega_{nk})
 |\mathbf{d}_{nk}^\bot|^2]
\end{eqnarray}
[$\mathbf{d}_{10}^\parallel\! =\! ((d_{10})_x,(d_{10})_y,0)$,
$\mathbf{d}_{10}^\perp\! =\! (0,0,(d_{10})_z)$].
Equation~(\ref{Ur2}) can be further investigated analytically in the
limits of short and long atom--interface separations. In the
nonretarded regime where $z_A\omega_{nk}/c\ll 1$  we can assume
$\beta\simeq \beta_1\simeq\mi q$ which leads to
\begin{equation}
r_s=\frac{\mu-1}{\mu+1}\,,\quad
r_p=\frac{\varepsilon-1}{\varepsilon+1}\,.
\end{equation}
Carrying out the integral in (\ref{Green}) gives 
\begin{eqnarray}
\label{Gnr} 
U_n(z_A)=\nonumber\\
-\sum_{k<n}\frac{|\mathbf{d}_{nk}^\parallel|^2
+2|\mathbf{d}_{nk}^\bot|^2}
{32\pi\varepsilon_0z_A^3}\,
\frac{|\varepsilon(\omega_{nk})|^2-1}
{|\varepsilon(\omega_{nk})+1|^2}
\end{eqnarray}
unless the half space is purely magnetic, in which case
\begin{eqnarray}
\label{Gnr2} 
U_n(z_A)=\nonumber\\
-\sum_{k<n}\frac{\mu_0\omega_{nk}^2|\mathbf{d}_{nk}^\parallel|^2}
{16\pi z_A}\,
\frac{|\mu(\omega_{nk})|^2-1}{|\mu(\omega_{nk})+1|^2}\,.
\end{eqnarray}
This shows that close to the surface, the resonant CP potential 
is attractive for $|\varepsilon|>1$ and ~--in the case of a purely
magnetic material--~ it is attractive for $|\mu|>1$. In particular,
for meta-materials with $|\varepsilon|<1$ and/or $|\mu|<1$, a
repulsive near-surface behavior of the potential can be expected. 

In the retarded regime $z_A\omega_{nk}/c\gg 1$, the main contribution
to the integral in Eq.~(\ref{Green}) is due to the stationary-phase
point $q=0$, so the reflection coefficients are approximated by
\begin{equation}
\label{r21r}
r_s=-r_p=\frac{\sqrt{\mu}-\sqrt{\varepsilon}}
{\sqrt{\mu}+\sqrt{\varepsilon}}\,.
\end{equation}
After substituting (\ref{r21r}) into (\ref{Green}), the integral can
be carried out and keeping only the leading order in
$c/(z_A\omega_{nk})$, the retarded CP potential~(\ref{Ur2}) takes the
form 
\begin{eqnarray}
\label{Uret}
U_n(z_A)&=&
 \sum_{k<n}\frac{\mu_0\omega_{nk}^2|\mathbf{d}_{nk}^\parallel|^2}
 {8\pi z_A}\nonumber\\
 &&\times\mathrm{Re}\Biggl\{e^{2\mi z_A\omega_{nk}/c}\,
 \frac{\sqrt{\varepsilon(\omega_{nk})}-\sqrt{\mu(\omega_{nk})}}
 {\sqrt{\varepsilon(\omega_{nk})}+\sqrt{\mu(\omega_{nk})}}\Biggr\}
 \nonumber\\
&=&\pm\sum_{k<n}\frac{\mu_0\omega_{nk}^2|\mathbf{d}_{nk}^\parallel|^2}
 {8\pi z_A}\,\cos(2z_A\omega_{nk}/c),
\end{eqnarray}
where the second equality holds for strongly electric and magnetic
half spaces, respectively. It can be seen that the retarded potential
is dominated by an oscillating term of decreasing amplitude where
purely electric and purely magnetic materials give rise to potentials
of different signs. 

\begin{figure}[t]
\begin{center}
\includegraphics[width=\linewidth]{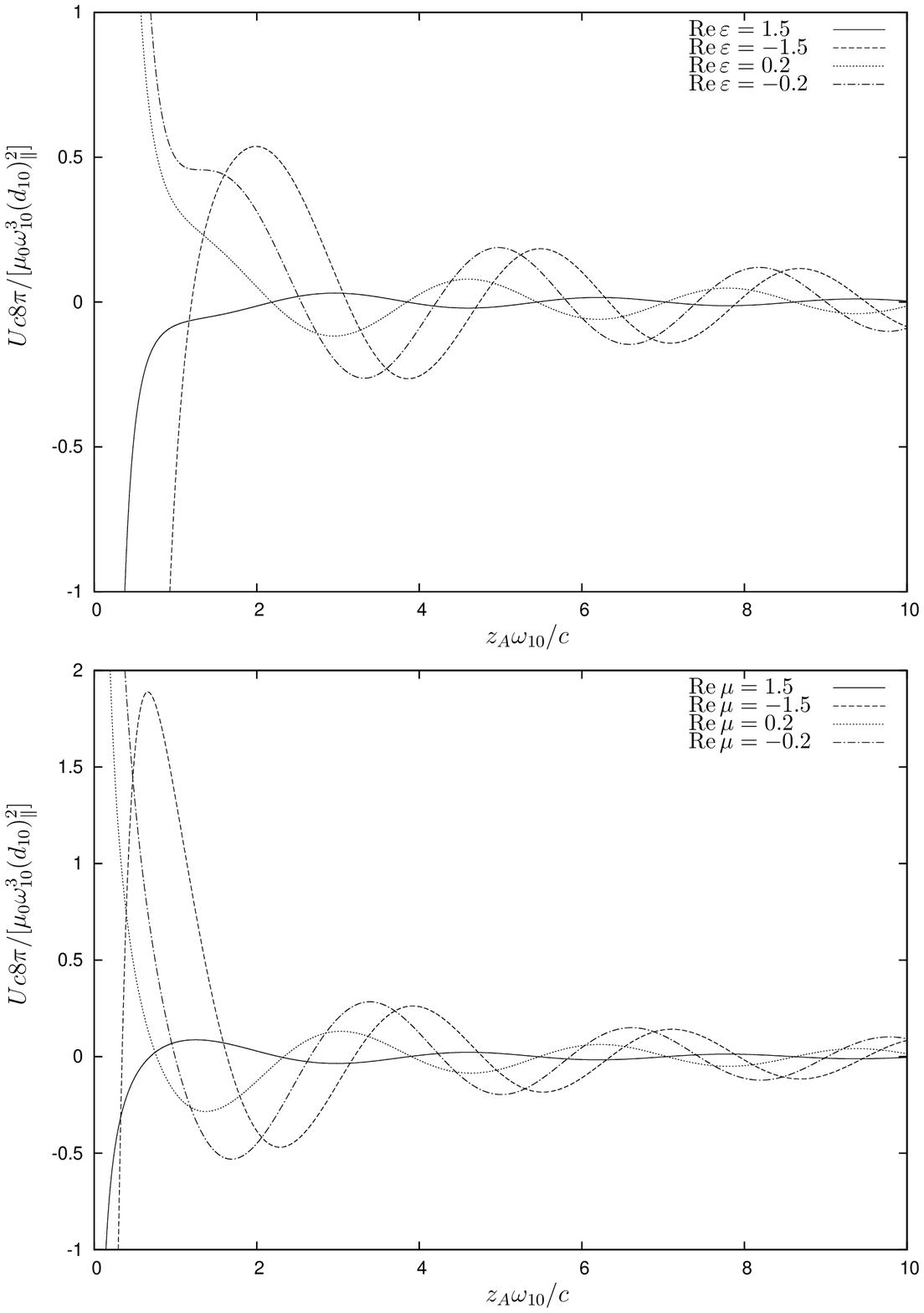}
\caption{\label{epsfig}
Resonant CP potential of a two-level atom in front of an
electric (above) and magnetic (below) meta-material half space for
different strengths of the electric/magnetic properties. 
The atomic dipole moment is oriented parallel to the surface and we
have assumed $\mathrm{Im}\epsilon=\mathrm{Im}\mu=10^{-3}$.
}
\end{center} 
\end{figure}

In Fig.~\ref{epsfig}, we show the resonant potential~(\ref{Ur2}) 
[together with Eqs.~(\ref{Green})--(\ref{beta})] for purely electric
(above) and purely magnetic (below) materials. As expected from the
retarded limit, the potential at large distances is governed by
decaying oscillations whose phase depends on the signs and
electric/magnetic nature of the medium response. The amplitude is of
these oscillations is largest for large negative
$\mathrm{Re}\varepsilon$ or $\mathrm{Re}\mu$. The figure confirms the
predicted attraction (repulsion) in the nonretarded regime for
$|\varepsilon|>(<)1,|\mu|>(<)1$ and further reveals that a repulsive
potential barrier can form for $\mathrm{Re}\varepsilon<-1$ or
$\mathrm{Re}\mu<-1$; in the latter case, it is even more pronounced.

\begin{figure}[t]
\begin{center}
\includegraphics[width=\linewidth]{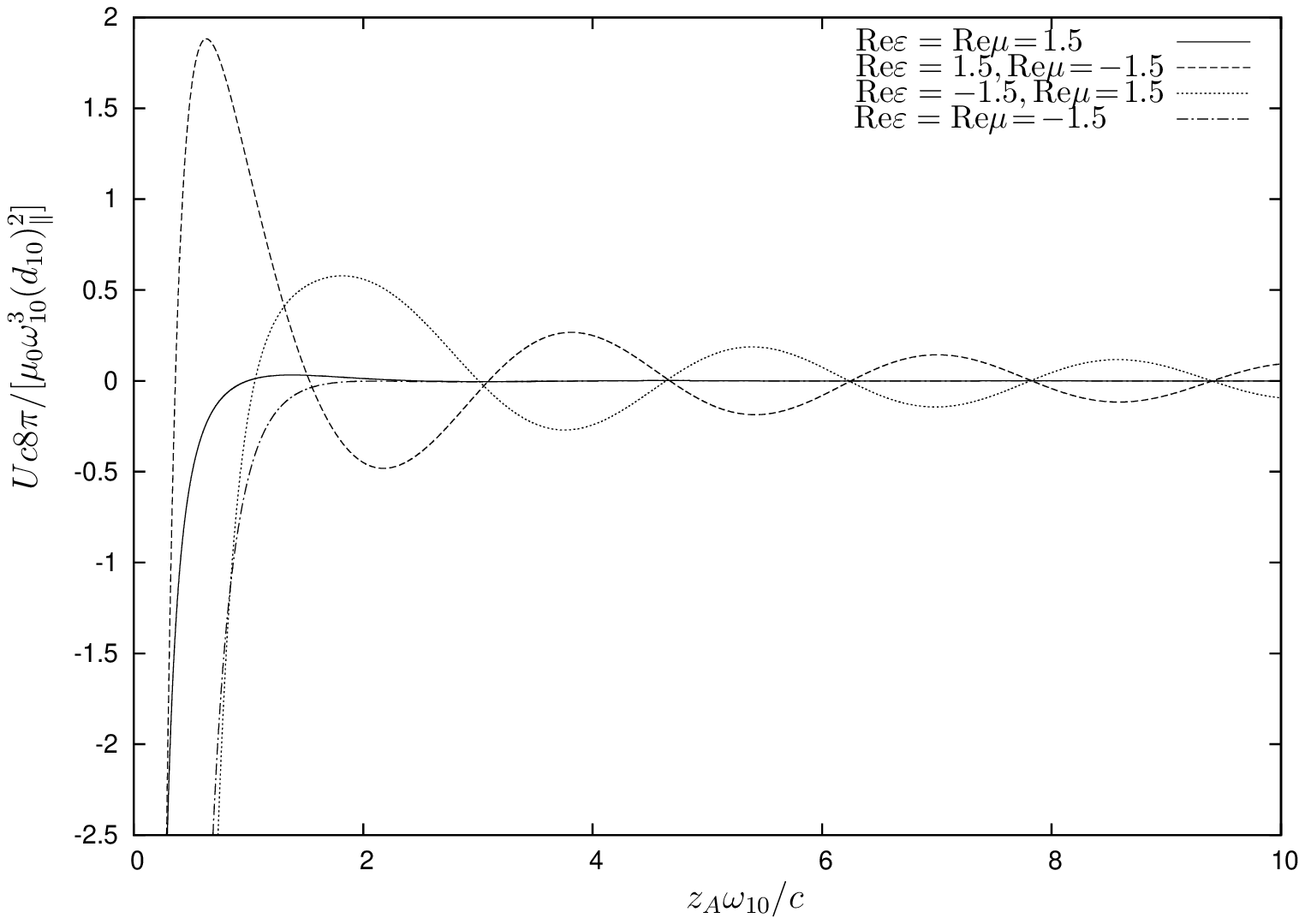}
\caption{\label{metafig}
Resonant CP potential of a two-level atom in front of a
magneto-electric
meta-material half-space. 
Assumptions of Fig.~\ref{epsfig} apply.
} 
\end{center}
\end{figure}

As a second example we have considered a genuinely magneto-electric
half space with different signs for $\mathrm{Re}\varepsilon$ and
$\mathrm{Re}\mu$ (see Fig.~\ref{metafig}). The strongest
oscillations are seen in the case of a meta-material with ${\rm Re}
\varepsilon<0$ and ${\rm Re} \mu >0$ (or vice versa), where the
oscillations have opposite sings in the two cases, as predicted from
Eq.~(\ref{Uret}).
The oscillation amplitude is very weak for a left-handed material or an
ordinary one with ${\rm Re} \varepsilon, {\rm Re} \mu>0$. 
%
%

\section{Perfect lens geometry}
\label{plg}
 
Consider an excited atom placed next to a meta-material superlens
(i.e., a plate of thickness $d$ with
$\mathrm{Re}\varepsilon=\mathrm{Re}\mu=-1$) bounded by a perfectly
conducting mirror on the far side (this arrangement was first
suggested in Ref.~\cite{Fleischhauer2005}). The reflection
coefficients of the associated Green tensor~(\ref{Green})
are given by
\begin{eqnarray}
\label{r2-s}
r_s=& \frac{\mu\beta-\beta_1-(\mu\beta+\beta_1)
e^{2\mi\beta_1d}}{\mu\beta+\beta_1-(\mu\beta-\beta_1)
e^{2\mi\beta_1d}}\,,\\
\label{r2-p}
r_p=&
\frac{\varepsilon\beta-\beta_1+(\varepsilon\beta+\beta_1)e^{
 2\mi\beta_1d}}
 {\varepsilon\beta+\beta_1+(\varepsilon\beta-\beta_1)
e^{2\mi\beta_1d}}\,.
\end{eqnarray}
We have discussed the associated resonant potential~(\ref{Ur2}) in
Ref.~\cite{Sambale2008a}. As shown, for a perfect, nonabsorbing lens,
the reflection coefficients reduce to $r_s=-r_p=-e^{-2\mi\beta d}$ and
the the potential~(\ref{equ1}) takes the form
\begin{eqnarray}
\label{equ20}
U(z_A)
&=-\sum _{k<n}\frac{\mu_0\omega_{nk}^3}{4\pi c \tilde{z}^3}
\left\{
 \left[\cos(\tilde{z})+\tilde{z}\sin(\tilde{z})\right.\right.\nonumber\\
&-\left.\tilde{z}^2\cos(\tilde{z})\right]
|\mathbf{d}_{nk}^\parallel|^2\nonumber\\
&+ \left.2
 \left[\cos(\tilde{z})+\tilde{z}\sin(\tilde{z})\right]
|\mathbf{d}_{nk}^\perp|^2
\right\}
\end{eqnarray}
[$\tilde{z}=2\omega_{k0}(z_A-d)/c$]. The atom is thus strongly
attracted to the focal plane of the lens at ($z_A=d$) where it
coincides with its image (which for $z_A>d$ is situated at $2d-z_A$). 

\begin{figure}[t]
\begin{center}
\includegraphics*[width=\linewidth]{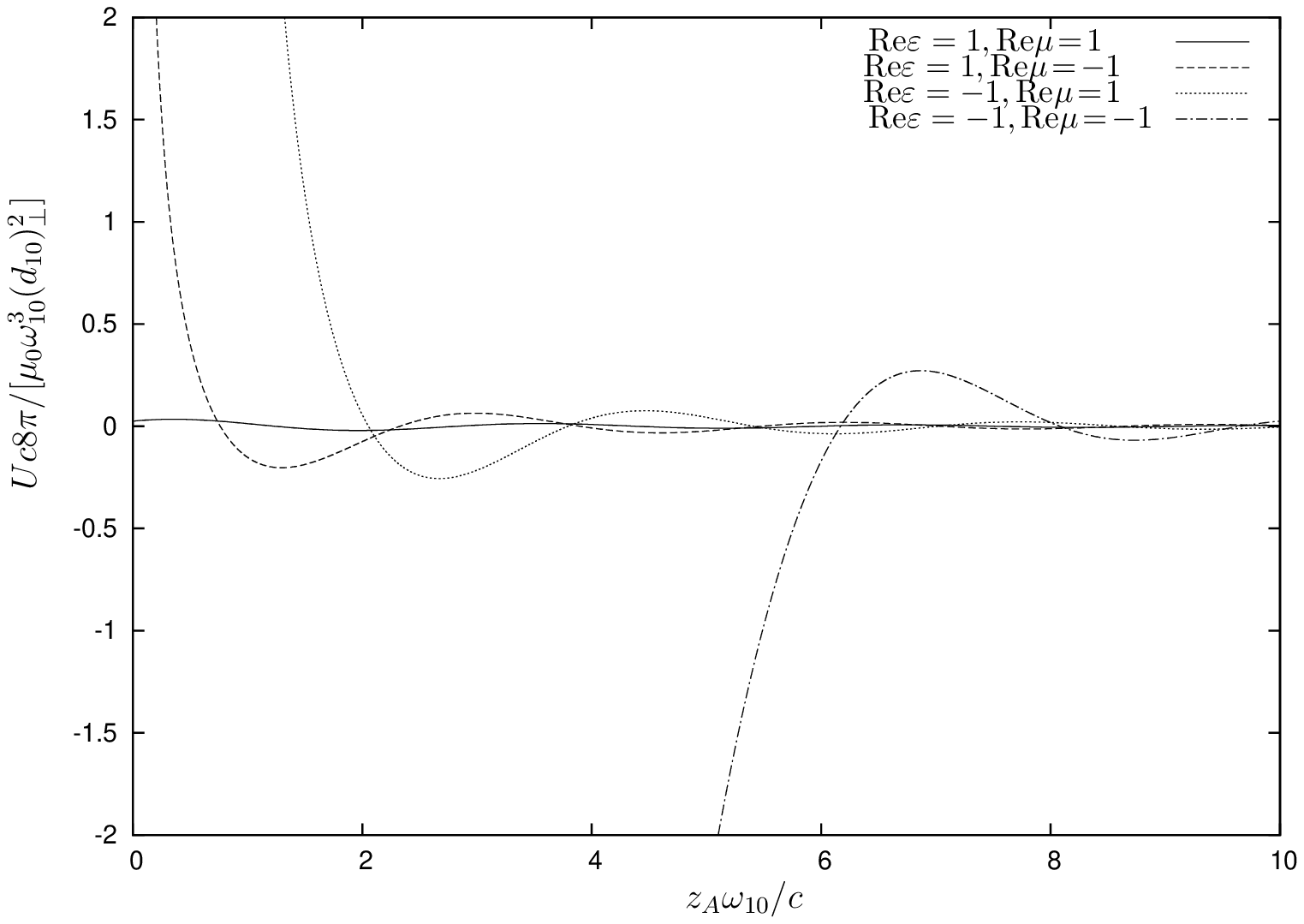}
\caption{Resonant CP potential of an excited two-level atom in front
of a meta-material slab of thickness $d=5c/\omega_{10}$ with a perfect
mirror at its far end. The atomic dipole moment is oriented
perpendicular to the surface and we have assumed
$\mathrm{Im}\epsilon=\mathrm{Im}\mu=10^{-4}$.
}
\end{center}
\label{fig2}
\end{figure}

For a more realistic absorbing lens, the potential no longer shows the
unphysical divergence at the focal plane, but an enhanced attraction
around the focal plane and thus away from the surface of the lens can
still be found for sufficiently weak absorption. Let us investigate
whether enhanced attraction away from physical surfaces is an effect
which is genuinely due to left-material properties. For that end, we
have calculated the CP potential~(\ref{Ur2}) [together with
Eqs.~(\ref{Green}), (\ref{beta}), (\ref{r2-s}) and (\ref{r2-p})] for a
two-level atom near a weakly absorbing superlens and compared it with
comparable righthanded meta-material slabs, with the results being
displayed in Fig.~\ref{fig2}. It is seen that the superlens gives rise
to enhanced attraction which sets in around the focal plane, while the
righthanded materials do not give rise to such a behavior. Instead,
the highly transparent material
$\mathrm{Re}\varepsilon=\mathrm{Re}\mu=1$ leads to a weak oscillating
potential which is entirely due to the mirror while the materials with
different signs of $\mathrm{Re}\varepsilon$ and $\mathrm{Re}\mu$ lead
to near-surface potential barriers. As seen in Sec.~\ref{rvdW}, the
latter behavior is due to the negative $\mathrm{Re}\varepsilon$ and
$\mathrm{Re}\mu$, respectively.
%
%
%


\section{Summary}

In this article we have studied the resonant CP potential of an
excited atom in front of meta-material half space. We have shown that
for long distances the potential exhibits attenuated oscillations,
while close to the surface the potential becomes attractive or
repulsive, depending on whether the absolute values of permittivity
and permeability are larger or smaller than unity.
 
Furthermore, we have reconsidered the potential of an excited atom
placed near a weakly absorbing left-handed planar superlens with a
perfect mirror at the far end. A comparison with comparable
righthanded meta-materials has shown that an enhanced superlens
attraction away from the surface is genuinely due to the left-handed
properties of the lens, while the potentials of righthanded materials
closely resemble those of respective half spaces without the
additional mirror.


\ack

The work was supported by Deutsche Forschungsgemeinschaft.
We acknowledge funding from the Alexander von Humboldt
Foundation (H.T.D. and S.Y.B.)
and the Vietnamese Basic Research Program (H.T.D.).


\end{document}